\documentclass{article}
\usepackage{spconf,amsmath,graphicx}
\usepackage{xcolor}
\usepackage{colortbl,booktabs}
\usepackage{graphicx}
\usepackage{xcolor}
\usepackage{fancyhdr}
\usepackage{tabularx}
\usepackage{multirow}
\usepackage{balance}
\usepackage{marvosym}

\usepackage{fancyhdr}

\title{MMGL: Multi-Scale Multi-View Global-Local Contrastive learning for Semi-supervised Cardiac Image Segmentation}
\name{\begin{tabular}{c}Ziyuan Zhao$^{\dag\ddag\sharp}$, Jinxuan Hu$^{\S}$, Zeng Zeng$^{\dag,\ddag}$\textsuperscript{,\Letter},  Xulei Yang$^{\dag, \ddag}$, Peisheng Qian$^{\dag}$\\
Bharadwaj Veeravalli$^{\S}$, Cuntai Guan$^{\ddag}$\end{tabular}}

\address{$^{\dag}$Institute of Infocomm Research (I$^2$R), A*STAR, Singapore\\
$^{\ddag}$ Artificial Intelligence, Analytics And Informatics (AI$^3$), A*STAR, Singapore\\
$^{\sharp}$School of Computer Science and Engineering, Nanyang Technological University, Singapore \\
$^{\S}$School of Electrical and Computer Engineering, National University of Singapore, Singapore 
}
\begin{document}
%
\maketitle

\thispagestyle{fancy}
\fancyhead{}
\lhead{}
\vspace{-0.5pt}
\lfoot{\footnotesize{Copyright 2022 IEEE. Published in 2022 IEEE International Conference on Image Processing (ICIP), scheduled for 16-19 October 2022 in Bordeaux, France. Personal use of this material is permitted. However, permission to reprint/republish this material for advertising or promotional purposes or for creating new collective works for resale or redistribution to servers or lists, or to reuse any copyrighted component of this work in other works, must be obtained from the IEEE. Contact: Manager, Copyrights and Permissions / IEEE Service Center / 445 Hoes Lane / P.O. Box 1331 / Piscataway, NJ 08855-1331, USA. Telephone: + Intl. 908-562-3966.}}
\cfoot{}
\rfoot{}

\begin{abstract}
With large-scale well-labeled datasets, deep learning has shown significant success in medical image segmentation. However, it is challenging to acquire abundant annotations in clinical practice due to extensive expertise requirements and costly labeling efforts. Recently, contrastive learning has shown a strong capacity for visual representation learning on unlabeled data, achieving impressive performance rivaling supervised learning in many domains. In this work, we propose a novel multi-scale multi-view global-local contrastive learning (MMGL) framework to thoroughly explore global and local features from different scales and views for robust contrastive learning performance, thereby improving segmentation performance with limited annotations. Extensive experiments on the MM-WHS dataset demonstrate the effectiveness of MMGL framework on semi-supervised cardiac image segmentation, outperforming the state-of-the-art contrastive learning methods by a large margin.
\end{abstract}

\textbf{\begin{figure*}[tp]
    \includegraphics[height=0.38\textwidth,width=0.77\textwidth]{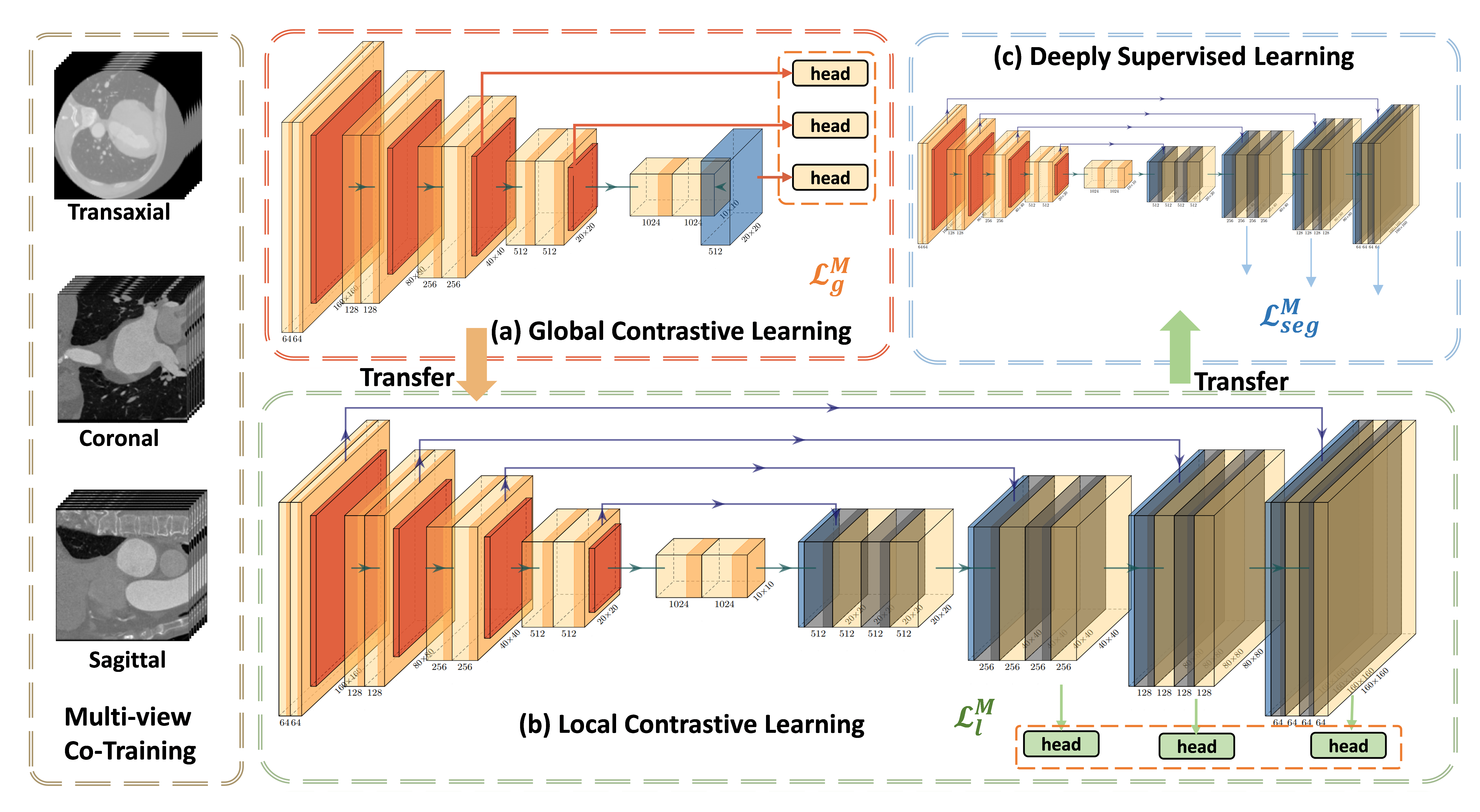}
    \centering
    \caption{Overview of the proposed multi-scale multi-view global-local contrastive learning~(MMGL)  framework, consisting of (a) Multi-scale global unsupervised contrastive learning, (b) Multi-scale local supervised contrastive learning, and (c) Multi-scale deeply supervised learning. The multi-view slices are provided for multi-view contrastive learning.}
    \label{fig:archi}
\end{figure*}}

\section{Introduction}
\label{sec:intro}
Cardiovascular disease is one of the most serious threats to human health globally~\cite{chen2020deep}, and accurate cardiac substructure segmentation can help disease diagnosis and treatment planning~\cite{savaashe2019review,lei2020medical}. To achieve the great success of deep learning methods on automatic cardiac image segmentation~\cite{oktay2017anatomically}, massive densely pixel-wise annotations are demanded for training, which are complicated to acquire in clinical practice since the annotation process is expensive, laborious, and time-consuming, leading to label scarcity. In this regard, many not-so-supervised methods, including active learning~\cite{zhao2021dsal,zhao2022self}, semi-supervised learning~\cite{bai2017semi,zhao2019semi} and self-supervised learning~\cite{zhou2021models,liu2021self} have been proposed to reduce annotation efforts. Recently, self-supervised learning~(SSL) has received much attention because of its powerful ability to learn image-level representations from unlabeled data, which is beneficial to various downstream tasks, such as classification~\cite{jing2020self}. As a successful variant of SSL, contrastive learning~\cite{jaiswal2021survey} focuses on transformation consistency on embedding latent features of similar images without labels, extracting transferable image-level representations for downstream tasks. Many contrastive learning methods have been proposed,~\emph{e.g.}, MoCo~\cite{he2020momentum} and SimCLR~\cite{chen2020simple}, achieving state-of-the-art performance on a variety of classification problems.

Most existing contrastive learning methods aim to realize image-level classification, whereas semantic segmentation requires pixel-wise classification~\cite{jaiswal2021survey}. Some recent work~\cite{chaitanya2020contrastive, zeng2021positional, hu2021semi} were developed to learn dense representations by introducing pixel-wise contrastive loss for medical image segmentation. Despite the success on medical image segmentation, most of them only applied contrastive learning on the last layer of encoder or decoder, ignoring the rich multi-scale information for robust representation learning. Besides, volumetric information is still underexplored for contrastive learning due to computational complexity. Intuitively, features of similar images extracted from the same specific model layer (\emph{e.g.}, the last encoder layer) should be similar, based on the principle of contrastive learning, and different views of the same anatomical structure can provide additional spacial contextual information~\cite{wei2019m3net,xia2018bridging}. These motivate us to extend contrastive learning into multiple layers at different scales and views, to expect extra benefits from multi-scale multi-view representations.

In this work, we propose a novel \textbf{M}ulti-scale \textbf{M}ulti-view \textbf{G}lobal-\textbf{L}ocal contrastive learning framework (MMGL), for semi-supervised cardiac image segmentation. In MMGL, we leverage multi-view multi-scale features from different layers and planes via contrastive learning to pre-train the segmentation network. Specifically, we first pre-train the encoder by applying global unsupervised contrastive learning to different encoder layers with unlabeled data. Next, we further pre-train the decoder by adopting local supervised contrastive learning with labeled data at different decoder layers. Finally, we inject deep supervision into the pre-trained segmentation model for fine-tuning with limited annotations. By combining the three strategies with multi-view co-training, our method can efficiently leverage both multi-view labeled and unlabeled data for semi-supervised cardiac image segmentation. Experimental results on Multi-Modality Whole Heart Segmentation Challenge 2017~(MM-WHS) dataset demonstrate that our approach is superior to the state-of-the-art baselines in the semi-supervised learning setting.

\section{Related Work}

Recent years have witnessed the powerful capacity of self-supervised learning~(SSL) on representation learning and downstream tasks~\cite{liu2021self}. By constructing unsupervised losses with unlabeled data for pre-training, useful representations can be learned and transferred to various downstream tasks, such as classification. More recently, contrastive learning~\cite{jaiswal2021survey}, a promising SSL variant, has received much attention, in which a contrastive loss is employed to encourage representations to be similar~(dissimilar) for similar~(dissimilar) pairs. The intuition behind this is that the augmented views of the same images should have similar representations, and views of the different images should have dissimilar representations. Following the philosophy, many works have been proposed for image-level representation learning~\cite{jaiswal2021survey}, such as SimCLR~\cite{chen2020simple}.

For pixel-level image segmentation, Chaitanya~\emph{et al.}~\cite{chaitanya2020contrastive} proposed a two-step contrastive learning strategy for medical image segmentation, in which both image-level global features and pixel-level local features were extracted from encoder and decoder, respectively, for follow-up fine-tuning on limited annotations. More recently, Hu~\emph{et al.}~\cite{hu2021semi} advanced the local contrastive learning with label information in the pre-training phase, showing its effectiveness on label-efficient medical image segmentation. However, these works focus on representation learning from single-view slices~\emph{e.g.}, coronal plane or from one single layer of networks,~\emph{e.g.}, final encoder layer, leaving rich multi-scale and multi-view information unexplored. On the other hand, multi-scale learning, which aims to learn discriminative feature representations at multiple scales, has been largely applied in various computer vision tasks~\cite{dou20163d,li2021hierarchical}. Multi-view learning also has been widely adopted to improve medical image segmentation performance~\cite{zhang2020exploring}. In this regard, we are motivated to advance contrastive learning into a multi-scale multi-view version for more efficient semi-supervised medical image segmentation.

\section{Methodology}
In Fig.~\ref{fig:archi}, we illustrate the proposed multi-scale multi-view global-local semi-supervised learning (MMGL) framework. We fully take advantage of multi-view co-training and multi-scale learning in both the pre-training and fine-tuning steps of the segmentation workflow. To be more specific, for the pre-training stage, we first design a multi-scale global unsupervised contrastive learning module to extract global features from different views and encoder layers. Next, we develop a multi-scale local supervised contrastive learning scheme by adding the decoder layers on top of the encoder to extract multi-scale local features from different views. In the fine-tuning stage, we train the segmentation model with multi-scale deeply supervised learning with a few labels.
%
\subsection{Multi-scale Global Unsupervised Contrastive Learning}
We adopt the U-Net architecture~\cite{ronneberger2015u} as our backbone network. To form contrastive pairs, we forward propagate a batch of multi-view inputs $\left\{ {x_{1}},{x_{2}},...,{x_{b}}\right\}$ through two random augmentation pipelines to get an augmentation set $A$ = $\left\{ {a_{1}},{a_{2}},...,{a_{2b}}\right\}$, where ${a_{i}}$ is an augmented input. To extract multi-scale global representations, we add projection heads $\mathit{h(\cdot)}$ after each block of the encoder $\mathit{E(\cdot)}$ and the resulting representation denotes as ${z_{i}=h(E(x_{i}))}$. Then, we calculate the global unsupervised loss at different layers. Formally, the global unsupervised contrastive loss~\cite{hu2021semi} of the $e$-th layer in the encoder is defined as:
\begin{equation}
\mathbf{\mathcal{L}}_{g}^{e}=-\frac{1}{\left| A\right|}\sum_{i  \in  I}log\frac{exp(sim(z_{i} , z_{j})/\tau ) }{\sum_{k\in I-{\left\{ j\right\}}}exp(sim(z_{i} ,  z_{k})/\tau )},\nonumber
\end{equation}
where ${z_{i}=h(E(x_{i}))}$ and ${z_{j}=h(E(x_{j}))}$ are two augmented representations of the same image, and they form a positive pair. While ${z_{i}}$ and ${z_{k}}$ $(k\neq j)$ form a negative pair, $e$ stands for the $e$-th block of the encoder, $I$ is the index set of $A$, $\tau$ denotes the temperature coefficient, and $sim(\cdot )$ is the cosine similarity measurement. The global contrastive loss can pull the similar pairs closer and push away the dissimilar ones~\cite{chen2020simple}. 

Then, we sum up global contrastive losses from different levels and define the multi-scale global unsupervised contrastive loss as:
\begin{equation}
\mathbf{\mathcal{L}}_{g}^{M}=\sum_{e\in E} \lambda _{g}^{e}{\mathcal{L}}_{g}^{e}, \nonumber
\end{equation}
where $\lambda _{g}^{e}$ is the balancing weight of ${\mathcal{L}}_{g}^{e}$.

\subsection{Multi-scale Local Supervised Contrastive Learning}
Multi-scale global unsupervised contrastive learning only learns the image-level features. To learn the dense pixel-level features, we introduce the local supervised contrastive loss~\cite{hu2021semi} for a feature map $f_{l}$:
\begin{equation}
\mathbf{\mathcal{L}}_{f}(x)=-\frac{1}{\left| \Omega \right|}\sum_{i\in \Omega }\frac{1}{\left| P(i)\right|}log\frac{\sum_{i_{p}\in P(i)}exp(f_{l}^{i}\cdot  f_{l}^{i_{p}}/\tau )}{\sum_{i_{n}\in N(i)}exp(f_{l}^{i}\cdot  f_{l}^{i_{n}}/\tau )}, \nonumber
\end{equation}
where $i$ is the $i$-th point in the extracted feature map $f_{l}$. $\Omega$ contains the total points in the $f_{l}$. $P(i)$ contains the positive points set of $f_{l}$ that share the same annotation and $N(i)$ contains the negative set. The local supervised contrastive loss is:
\begin{equation}
\mathbf{\mathcal{L}}_{l}^{d}=\frac{1}{\left| A\right|}\sum_{a_{j}\in{A}}\mathbf{\mathcal{L}}_{f}(a_{j}), \nonumber
\end{equation}
where $d$ denotes the $d$-th upsampling block of the decoder~$\mathit{D(\cdot)}$. The multi-scale local supervised contrastive loss is defined as:
\begin{equation}
\mathbf{\mathcal{L}}_{l}^{M}=\sum_{d\in D} \lambda _{l}^{d}{\mathcal{L}}_{l}^{d}, \nonumber
\end{equation}
where $\lambda _{l}^{d}$ is the balancing weight of ${\mathcal{L}}_{l}^{d}$.

\subsection{Multi-scale Deeply Supervised Learning}
In the last fine-tuning stage, we adopt the multi-scale deep supervision strategy with a small portion of labeled data only in the transaxial view. The deeply supervised loss is formulated as:
\begin{equation}
\mathbf{\mathcal{L}}_{seg}^{M}=\sum_{d\in D} \lambda_{dice}^{d} {\mathcal{L}}_{dice}^{d},\nonumber
\end{equation}
where ${\mathcal{L}}_{dice}$ is the Dice loss~\cite{hu2021semi}, $\lambda_{dice}^{d}$ is the balancing weight of ${\mathcal{L}}_{dice}^{d}$.
\section{Experiments}
\subsection{Dataset and Experimental Settings}
We evaluated the proposed MMGL framework on a public medical image segmentation dataset from MICCAI 2017 challenge,~\emph{i.e.}, MM-WHS (Multi-Modality Whole Heart Segmentation). We used the $20$ cardiac CT volumes in our experiments. The expert labels for MM-WHS consist of $7$ cardiac structures: left ventricle~(LV), left atrium~(LA), right ventricle~(RV), right atrium~(RA), myocardium~(MYO), ascending aorata~(AA), and pulmonary artery~(PA). The dataset was randomly split into the training set, validation set, and testing set in the ratio of $2:1:1$. For pre-processing, slices from the transaxial, coronal and sagittal planes were extracted, and the transaxial view is set as the main view for evaluation. All slices were resized to $160\times 160$, followed by min-max normalization. Three common segmentation metrics~\cite{oktay2017anatomically},~\emph{i.e.}, dice similarity coefficient~(DSC), pixel accuracy~(PixAcc), and mIoU were applied for performance assessment, and higher values would imply better segmentation performance.

U-Net~\cite{ronneberger2015u} was employed as the backbone, which contains $4$ downsampling blocks (encoders) and $4$ upsampling blocks (decoders). To obtain multi-scale features, we add three heads for each stage. More specifically, for multi-scale global contrastive learning, a project head was added on top of each encoder block, consisting of an average pooling layer followed by two linear layers to output one $128$-dimensional embedding. For multi-scale local contrastive learning, we added one project head on top of each decoder block consisting of two ${1\times1}$ convolutional layers. The extracted feature maps were downsampled with fixed stride $4$ to reduce computation complexity. The augmentation pipeline includes brightness transform, Gamma transform, Gaussian noise transform, and spatial transforms, like rotation and crop. The temperature $\tau$ is fixed to $0.07$ for both global and local contrastive learning. The weights $\lambda_g^e$ and $\lambda_l^d$ were empirically set $\left\{ 0.2,0.2,0.6\right\}$ and $\left\{ 0.2,0.2,0.6\right\}$, respectively.

\begin{table}[t]
\centering
\caption{Segmentation results of different approaches. The triple ($\lambda_{dice}^1$,~$\lambda_{dice}^2$,~$\lambda_{dice}^3$) after MMGL is a portfolio of weights in the Dice loss function.}
\label{tab:results2}
\setlength\tabcolsep{1.6pt}
\scalebox{0.73}{

\begin{tabular}{c|ccc|ccc|lcc} 
\hline
\multirow{3}{*}{Method}           & \multicolumn{9}{c}{\% labeled data in train set}                                                                                                                                                                                \\ 
\cline{2-10}
                                  & \multicolumn{3}{c|}{10\%}                                           & \multicolumn{3}{c|}{20\%}                                                 & \multicolumn{3}{c}{40\%}                                                      \\ 
\cline{2-10}
                                  & Dice                 & mIoU                 & PixAcc                & Dice                 & mIoU                 & \multicolumn{1}{l|}{PixAcc} & \multicolumn{1}{c}{Dice} & mIoU                 & \multicolumn{1}{l}{PixAcc}  \\ 
\hline
Random                            & 0.569                & 0.422                & 0.959                 & 0.723                & 0.576                & 0.965                       & 0.817                    & 0.701                & 0.982                       \\ 
\hline
Global~                           & 0.606                & 0.455                & 0.964                 & 0.737                & 0.600                & 0.974                       & 0.829                    & 0.717                & 0.983                       \\

Global+Local~                     & 0.651                & 0.499                & 0.967                 & 0.787                & 0.660                & 0.974                       & 0.827                    & 0.716                & 0.983                       \\
SemiContrast~                      & 0.690                & 0.546                & 0.964                 & 0.799                & 0.678                & 0.974                       & 0.822                    & 0.712                & 0.983                       \\ 
\hline
MMGL (2:2:6)                      & 0.710                     & 0.566                     & 0.970                       & 0.815                     & 0.697                     & 0.978                            &  0.846    &  0.740                    &  0.985                           \\
MMGL (1:2:7)                      & 0.718                     & 0.567                    & 0.972                       & 0.813                     & 0.692                     & 0.977                            &  $\mathbf{0.849}$    &  $\mathbf{0.746}$                    &  $\mathbf{0.985}$                           \\
MMGL (3:3:4)                      & 0.739                     & 0.596                     & 0.974                       & $\mathbf{0.828}$                     & $\mathbf{0.714}$                     & $\mathbf{0.979}$                            &  0.844    &  0.740                    &  0.985                           \\
MMGL (2:3:5)                      & $\mathbf{0.743}$                     &$\mathbf{0.601}$                      & $\mathbf{0.973}$                       & 0.826                     & 0.712                     & 0.979                            &  0.848    &  0.742                    &  0.985                           \\
\hline
\end{tabular}
}

\end{table}

\subsection{Baseline Methods}
Besides the random initialization baseline, we compared the proposed MMGL framework with various contrastive learning methods, including Global~\cite{chen2020simple}, Global+Local~\cite{chaitanya2020contrastive}, and SemiContrast~\cite{hu2021semi}. Specifically, Global~\cite{chen2020simple} is the SimCLR framework, which pre-trains the encoder via contrastive learning on unlabeled data, while Global+Local~\cite{chaitanya2020contrastive} pre-trains both the encoder and decoder. Furthermore, SemiContrast~\cite{hu2021semi} introduces label information for local contrastive learning.




\subsection{Experimental Results and Discussions}
\subsubsection{Comparison with existing methods}
\begin{figure}[!thb]
\centering
\includegraphics[width=0.47\textwidth]{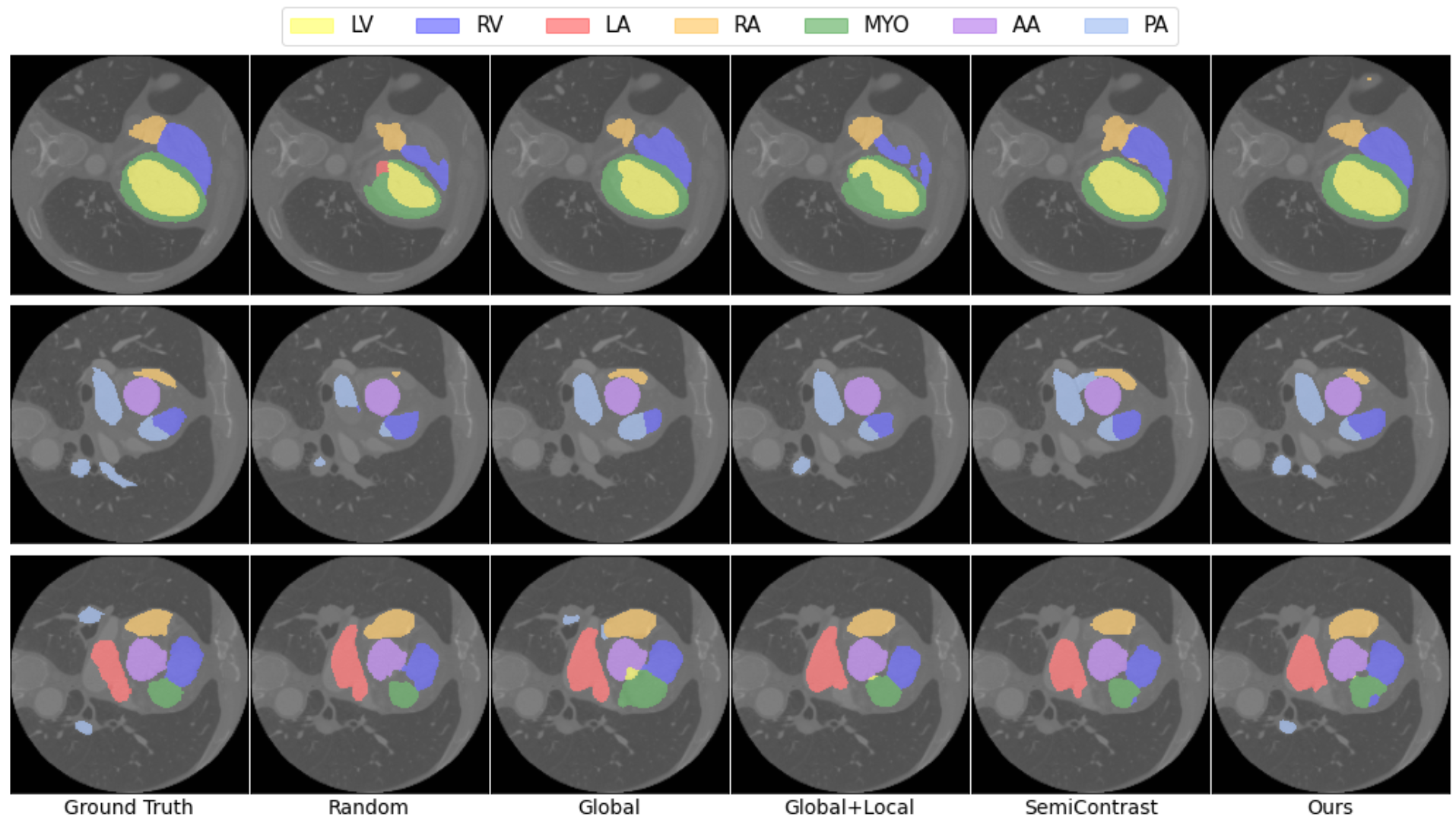}
\centering
\caption{A visual comparison of segmentation results produced by different methods.}
\label{fig:visual}
\end{figure}

The experimental results are shown in Table~\ref{tab:results2}. Compared to the Random baseline, all different contrastive learning methods can improve the performance against label scarcity. Moreover, SemiContrast obtains better performance than others in most cases because it leverages label information for contrastive learning. It is noted that MMGL achieves up to $84.9\%$ in Dice score when training with $40\%$ labeled data, which is superior to other methods. We also observe that the proposed multi-scale multi-view contrastive learning method can substantially improve the segmentation performance with the highest values of all three metrics using different portions of labeled training data, demonstrating the effectiveness of multi-scale features for contrastive learning with multi-view images. Furthermore, we perform MMGL with varying portfolios of weights $\lambda_{dice}$ at the fine-tuning stage and observe that all combinations outperform other contrastive learning methods, despite some changes in performance. Fig.~\ref{fig:visual} demonstrates the visualization results of different methods, in which our method generates more accurate masks with clear boundaries and fewer false positives on different substructures.
\subsubsection{Ablation analysis of key components}
\begin{figure}[!thb]
\centering
\includegraphics[width=0.45\textwidth]{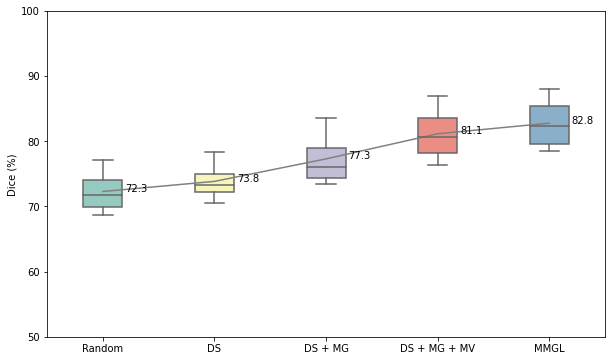}
\centering
\caption{Ablation analysis of key components of MMGL.}
\label{fig:ablation}
\end{figure}

To further analyze the effectiveness of different key components of MMGL, we perform a comprehensive ablation analysis (repeat $4$ times for mean and standard deviation), as shown in Fig.~\ref{fig:ablation}. Specifically, we start with the random baseline and achieve $72.3\%$ in Dice with $20\%$ labeled data. By adding deeply supervised learning~(DS), the metric is improved to $73.8 \%$. By adding multi-scale global learning~(MG), the metric is improved to $77.3 \%$. When applying multi-view~(MV) images for contrastive learning, the metric is enhanced by $3.8 \%$. Finally, MMGL increases the performance to $82.8 \%$. In total, the proposed method improves the performance by $10.5 \%$. These experiments demonstrate the effectiveness of the four components.

\section{Conclusions}
In this work, we propose a novel multi-scale multi-view global-local contrastive learning approach for semi-supervised medical image segmentation. We encourage more useful global and local features via multi-scale deep supervision from different encoders and decoder layers with multi-view images for contrastive learning and fine-tuning. Experimental results on the MM-WHS dataset demonstrate that our framework is superior to other state-of-the-art semi-supervised contrastive learning methods. The proposed MMGL method is simple yet effective for semi-supervised segmentation, which can be widely implanted for other segmentation networks and tasks.

\clearpage
\balance
\bibliographystyle{IEEEbib}
\bibliography{refs1}

\end{document}